\documentclass[pdftex,twocolumn,epjc3]{svjour3}
\usepackage[utf8]{inputenc}
\usepackage{amsmath}
\usepackage{amsfonts}
\usepackage{amssymb}
\usepackage{graphicx}

\DeclareMathOperator{\csch}{csch}
\DeclareMathOperator{\arcsinh}{arcsinh}
\usepackage{mathrsfs}
\usepackage{latexsym}
\usepackage{bm}
\usepackage{subfigure}
\usepackage{color}
\usepackage{hyperref}
\usepackage{subfigure}
\RequirePackage{flushend}

\usepackage[left=2cm,right=2cm,top=2cm,bottom=2cm]{geometry}

\journalname{Eur. Phys. J. C}
\begin{document}
\title{Signature flip in deceleration parameter: A thermodynamic phase transition?}
\author{Tanima Duary\thanksref{e1,addr1}
\and Narayan Banerjee\thanksref{e2,addr1}
\and Ananda Dasgupta\thanksref{e3,addr1}
}

\thankstext{e1}{e-mail: td14ip021@iiserkol.ac.in}
\thankstext{e2}{e-mail: narayan@iiserkol.ac.in}
\thankstext{e3}{e-mail: adg@iiserkol.ac.in}
\institute{Department of Physical Sciences, Indian Institute of Science Education and Research Kolkata,
Mohanpur 741 246 ,WB, India\label{addr1}}

\date{Received: date / Accepted: date}
\maketitle

\abstract{
Using the Hayward-Kodama temperature for the apparent horizon, it is found that matter content in the Universe is not thermodynamically stable, and the entry to the late accelerated expansion is actually a 
second order phase transition. The cosmological model used for the purpose is one that imitates the $\Lambda$CDM model, the favoured model for the present Universe.
}

\maketitle

\section{Introduction:}
The idea of a horizon and the thermodynamical properties of cosmological models were brought into being inspired by black hole thermodynamics\cite{Wald:1984rg,Poisson:2009pwt}.
This served the purpose as a diagnostics of the validity of cosmological models, mainly 
by checking the validity of the ``Generalized Second Law of Thermodynamics" (GSL)\cite{bekenstein1974generalized} in the model. The GSL states that the total entropy of the Universe is non-decreasing. For a comprehensive review, we refer to the monograph by 
Faraoni\cite{faraoni2015horizons}. This helps one to determine the favoured model amongst two or more. For example, we refer to the recent work which indicates that freezing models are better in this respect than the thawing models\cite{duary2019thawing}.

 Understanding the thermodynamic stability of the  matter content of the Universe offers insights into the underlying mechanisms governing cosmic processes. It allows us to discern the conditions under which the matter components maintain equilibrium or undergo transformative phases. This knowledge contributes to a deeper comprehension of how the cosmos evolves and goes through different phases.
Furthermore, the connection between thermodynamic stability and late accelerated expansion provides a unique vantage point for verifying and refining existing cosmological models. If such a correlation is confirmed, it implies a previously unnoticed interplay between thermodynamics and the evolution of the Universe. This can lead to novel perspectives on the driving forces behind the cosmic expansion and the role of various energy components. If thermodynamic stability is indeed linked to the accelerated expansion phase, it could offer a new perspective on the nature and behavior of dark energy, potentially guiding future research and theoretical developments in this realm. This correlation offers a means to select favoured propositions from various existing theories concerning dark energy in the literature. Thermodynamics is a robust branch of physics and survived all attempts at empirical falsification. It can establish bounds on physical processes.

 Recently, Barboza et al. \cite{Barboza:2015bha}, delved into the thermodynamic aspects of DE fluids. Their study encompasses both thermal and mechanical stability, necessitating positive values for heat capacities and fluid compressibility. Their findings indicated that the stability of DE fluid implies a negative constant equation of state parameter. However, their analysis encountered a contradiction with observational constraints set by type Ia supernovae, Baryon Acoustic Oscillations (BAO), and Hubble parameter data on a generalized DE fluid. Consequently, their conclusion suggested that DE fluid models might be regarded as unfeasible from a thermodynamic standpoint. 
On the other hand, as indicated in reference\cite{luongo2014cosmographic}, in order to achieve a universe currently undergoing an accelerated expansion, it becomes necessary to posit a negative value for specific heat at constant volume($C_V$). Their study further reveals that the specific heat at constant volume, $C_V$, exhibits a negative value in the present epoch, while the specific heat at constant pressure($C_P$), approaches a value close to zero. These outcomes are consistent with the $\Lambda$CDM model.

The thermodynamic stability of model can be ascertained from the properties of the second order derivatives of the internal entropy of the system. For some examples, see references \cite{Ferreira:2015iaa} and \cite{mukherjee2021nonparametric}. The motivation of 
the present work, to begin with, is to look at the thermodynamic stability of a cosmological model that mimics a $\Lambda$CDM model for the late time evolution. The method is to look for the concavity of the entropy function for the matter 
content in the Universe. This is done through the properties of the hessian matrix, which involves the second order derivatives of the entropy. An example of such treatments in cosmology can be found in the work by Bhandari, Haldar and Chakraborty\cite{bhandari2017interacting}.

With a metric ansatz where the scale factor is a hyperbolic function ($a\sim {\sinh}^{2/3}(t/t_0)$), that imitates a $\Lambda$CDM behaviour, it is found that the system is not thermodynamically stable. 

The calculation yields a surprising result. It is found that although the entropy $S$ is continuous, $C_V$, the thermal capacity at constant volume has a discontinuity at a particular value of the redshift $z$, where the evolution transits from 
the decelerated to the accelerated state of expansion. Clearly, this transition is a phase transition as the discontinuity is in $C_V$. It is also found that the deceleration parameter $q$ plays the role of the order parameter, and the 
order of the discontinuity is simply unity.
 
Unlike the case of a stationary black hole, the horizon in cosmology is evolving, and is defined as an apparent horizon. This motivates one to replace the Hawking temperature by Hayward-Kodama
 temperature \cite{hayward1998unified,hayward2009local,DiCriscienzo:2009kun} 
as the temperature of the horizon. This turns out to be a crucial difference. If the Hawking temperature is used instead, this phase transition goes missing! This may be the reason behind this important connection between the 
signature flip in $q$ and a second order phase transition. It deserves mention that some recent research on evolving black holes embedded in a de-Sitter spacetime, a second order phase transition is 
indicated\cite{Zhang:2022aqg,Zhao2018,ali2022thermodynamics,Zhang:2020odg,Ma:2018hni}. 

The paper is organized as follows. Section 2 introduces the general thermodynamic stability conditions. In section 3, the cosmological model  used in the work is discussed.  Section 4 deals with the stability analysis and the phase transition 
in the model. The 5th and final section includes some concluding remarks.

\section{General stability condition:}
We consider a spatially flat, homogeneous and isotropic universe given by the metric,
\begin{equation}
ds^2=-dt^2+a^2(t)[dr^2+r^2 d\Omega^2],
\end{equation}
where $a=a(t)$ is the scale factor. The corresponding Einstein field equations are, 
\begin{align} \label{fe}
3H^2 &=  \rho , \\
2\dot{H}&= - (\rho+p),
\end{align}
where $\rho$ and $p$ are the total energy density and pressure of the matter content, $H=\frac{\dot a}{a}$ is the Hubble parameter and an overhead dot indicates a derivative with respect to the cosmic time $t$. 
Units are chosen where $c=1$ and $8\pi G=1$. Radius of apparent horizon, $\tilde {r}_h$, defined as $g^{\mu\nu} \tilde{r}_{h_{,\mu}} \tilde{r}_{h_{,\nu}}=0$ , is ${\tilde {r}_h}=\frac{1}{H}$ for a spatially flat ($k=0$) 
universe\cite{faraoni2015horizons}.

We consider that the fluid inside the horizon is in thermodynamical equilibrium with the horizon. 
As shown by Mimoso and Pav\'{o}n\cite{mimosodiego2016}, it has been determined that the attainment of thermal equilibrium between radiation and the cosmic horizon is not possible. This is due to Wien's law, which consistently results in a wavelength greater than the horizon radius throughout all time periods. However, nonrelativistic particles can achieve equilibrium at a certain juncture in the expansion, contingent upon the mass of the particles. In this context, the conjecture posited by various researchers (such as in \cite{IZQUIERDO2006420},\cite{Setare_2006},\cite{Gong:2006ma}), thermal equilibrium between dark energy and the horizon is not without its rationale. In our study, we have not taken into account any radiation component. As a result, the assumption to consider thermodynamic equilibrium between the horizon and the fluid content remains applicable. 

 The temperature associated with a horizon is linked to its surface gravity through the equation $T = \frac{\kappa}{2 \pi}$, where $\kappa$ denotes the surface gravity. When dealing with a spacetime that remains unchanging, an event horizon falls under the category of a Killing horizon. The surface gravity ($\kappa$) is defined in relation to the Killing vector ($\xi^a$) using the equation:
\begin{equation}\label{sgravstatic}
\xi^a\nabla_a\xi^b = \kappa\xi^b.
\end{equation}
However, in cases characterized by dynamic conditions, the aforementioned concept encounters limitations due to the absence of a timelike Killing vector. Yet, Hayward introduced an alternative definition of surface gravity that applies to dynamic, spherically symmetric spacetimes. This new definition relies on the Kodama vector denoted as $K^a$. The Kodama vector is defined as\cite{Kodama:1979vn}, 
\begin{equation}
K^a \equiv \epsilon^{ab} \nabla_b R,
\end{equation}
where $R$ is the areal radius of the 2-sphere and $\epsilon^{ab}$ is the volume form of the 2-metric $h^{ab}$\cite{Wald:1984rg}. 
It can be expressed as follows:
\begin{equation}\label{HK surgrav}
\frac{1}{2}g^{ab}K^c(\nabla_c K_a-\nabla_a K_c)= \kappa_{\mbox{ko}} K^b.
\end{equation}
In this context, $\kappa_{\mbox{ko}}$ represents the surface gravity. For an extensive and detailed exploration of this concept, we recommend referring to the monograph authored by Faraoni\cite{faraoni2015horizons} (see also \cite{Duary:2022dta}).
Within the framework of a spatially flat FRW cosmology, the surface gravity associated with the apparent horizon is given by\cite{Cai}:
\begin{equation}
\kappa_{\mbox{ko}} = -\frac{1}{2H}(\dot{H}+2H^2).
\end{equation}
Consequently, building on the preceding analysis, one can find the Hayward-Kodama temperature\cite{hayward1998unified,hayward2009local,DiCriscienzo:2009kun} for the apparent horizon as,
\begin{align}\label{temp}
T &= \frac{\mid{\kappa_{\mbox{ko}}\mid}}{2\pi} \nonumber\\
&= \frac{2H^2+\dot{H}}{4\pi H}.
\end{align}

It deserves mention that the temperature vanishes if the scale factor has the form $a(t)=\sqrt{\alpha t^2+\beta t+\gamma}$. Therefore in a pure radiation dominated era, equation(\ref{temp}) yields a zero temperature. However, the present work is quite safe in this respect as it does not deal with radiation in any way.
The entropy of the horizon is 
\begin{equation}
\label{hor-entr}
S_h=2\pi A,
\end{equation}
where $A=4\pi \tilde{r}_h^2$ is the area of apparent horizon\cite{faraoni2015horizons}.

Differentiating $S_h$ with respect to time $t$, one obtains,
\begin{equation}
\dot{S_h}= -16\pi^2\frac{\dot{H}}{H^3}.
\end{equation}

For the fluid inside the horizon, first law of thermodynamics applied to a hydrostatic system  looks like,  
\begin{equation}\label{Gibbs}
TdS_{\mbox{in}}=dU+pdV,
\end{equation}
where $S_{\mbox{in}}$, $U$ and $V$ denote the entropy, the internal energy and the volume of the fluid inside the horizon respectively. $V$ is bounded by the apparent horizon, 
\begin{align}\label{vol}
V &=\frac{4}{3}\pi \tilde{r}_h^3 \nonumber \\ 
&=\frac{4}{3}\pi \frac{1}{H^3}
\end{align}

Rate of change of entropy of fluid inside the horizon is,
\begin{align}\label{ent}
\dot{S}_{\mbox{in}}&= \frac{1}{T_h}\left[(\rho+p)\dot{V}+\dot{\rho}V\right]\nonumber\\
&= \frac{1}{T_h}(\rho+p)(\dot{V}-3HV).
\end{align}
Now inserting $T_h$ from equation \eqref{temp} and $V$ from equation \eqref{vol} in \eqref{ent}, one obtains the expression of $\dot{S}_{\mbox{in}}$ as,
\begin{align}\label{ent-rate1}
\dot{S}_{\mbox{in}}  = 16\pi^2\frac{\dot{H}}{H^3}\left(1+\frac{\dot{H}}{2H^2+\dot{H}} \right).
\end{align}

Therefore rate of change of the total entropy is,
\begin{align}\label{ent-rate2}
\dot{S}&=\dot{S}_h+\dot{S}_{\mbox{in}} \nonumber\\
&= 16\pi^2\frac{\dot{H}^2}{H^3}\left(\frac{1}{2H^2+\dot{H}}\right).
\end{align}

\section{A model that mimics $\Lambda$CDM}

We assume a simple ansatz for the scale factor: 
\begin{equation}
\label{ansatz-a}
 \frac{a}{a_0}\sim\frac{\sinh^{2/3}(t/t_0)}{\sinh^{2/3}(1)},
\end{equation} which gives an accelerated expansion for a late time whereas as a decelerated expansion in the early matter dominated era. Here we have considered $a=a_0$ at $t=t_0$ and taken $t_0=1$. 
 It's worth highlighting that the chosen ansatz has been tailored to imitate the characteristics inherent in the $\Lambda$CDM model, the favoured model for the present Universe\cite{PADMANABHAN2003235}. In the era dominated by matter, the scale factor adheres to a behavior of $a(t)\propto t^{2/3}$. During the phase primarily influenced by dark energy, with the assumption of $w=-1$, the scale factor asymptotically follows a trend of $a(t)\propto \exp(Ht)$. When considering a Universe that maintains a spatially flat geometry and encompasses both matter and vacuum energy, the solution encompasses the attributes of these two components across early and late time periods. This comprehensive solution is expressed as, 
 \begin{align}
 a(t)&=\left(\frac{\Omega_m}{\Omega_{vac}}\right)^{1/3} \left(\sinh[3\sqrt{\Omega_{vac}}H_0 t/2]\right)^{2/3}\nonumber \\&=\left(\frac{\Omega_m}{\Omega_{vac}}\right)^{1/3}\sinh^{2/3}(t/t_0),
\end{align} effectively capturing the tendencies exhibited by matter and vacuum energy, while aligning with their behaviors as time progresses from earlier to later epochs and is essentially offers a $\Lambda$CDM model\cite{Frieman:2008sn}. In our study, we have included the sine hyperbolic behavior dependent on time into the scale factor. This inclusion enables us to replicate the intrinsic characteristics of the $\Lambda$CDM model while simplifying mathematical details.
  The equation \eqref{ansatz-a} can be used to write $t/t_0 = \arcsinh\left((\frac{1}{1+z})^{3/2}\sinh(1)\right)$, where $z$ is the redshift, defined as $1+z = \frac{a_0}{a}$, where $a_0$ is the present value of 
the scale factor.
One can write Hubble parameter in terms of $z$ as, 
\begin{align}
H &=\frac{2}{3}\coth (t/t_0)\nonumber\\
 &=\frac{2\csch(1)}{3}\frac{\sqrt{1+\frac{\sinh^2(1)}{(1+z)^3}}}{\left(\frac{1}{1+z}\right)^{3/2}}.
\end{align}
 Deceleration parameter, defined as $q=-\left[1+\frac{\dot{H}}{H^2}\right]$, looks like,
  \begin{equation}\label{q}
  q(z)= -1+\frac{3}{2\left(1+\frac{\sinh^2(1)}{(1+z)^3}\right)},
\end{equation}    
in terms of $z$.

\begin{figure}
\boxed{\includegraphics[scale=0.6]{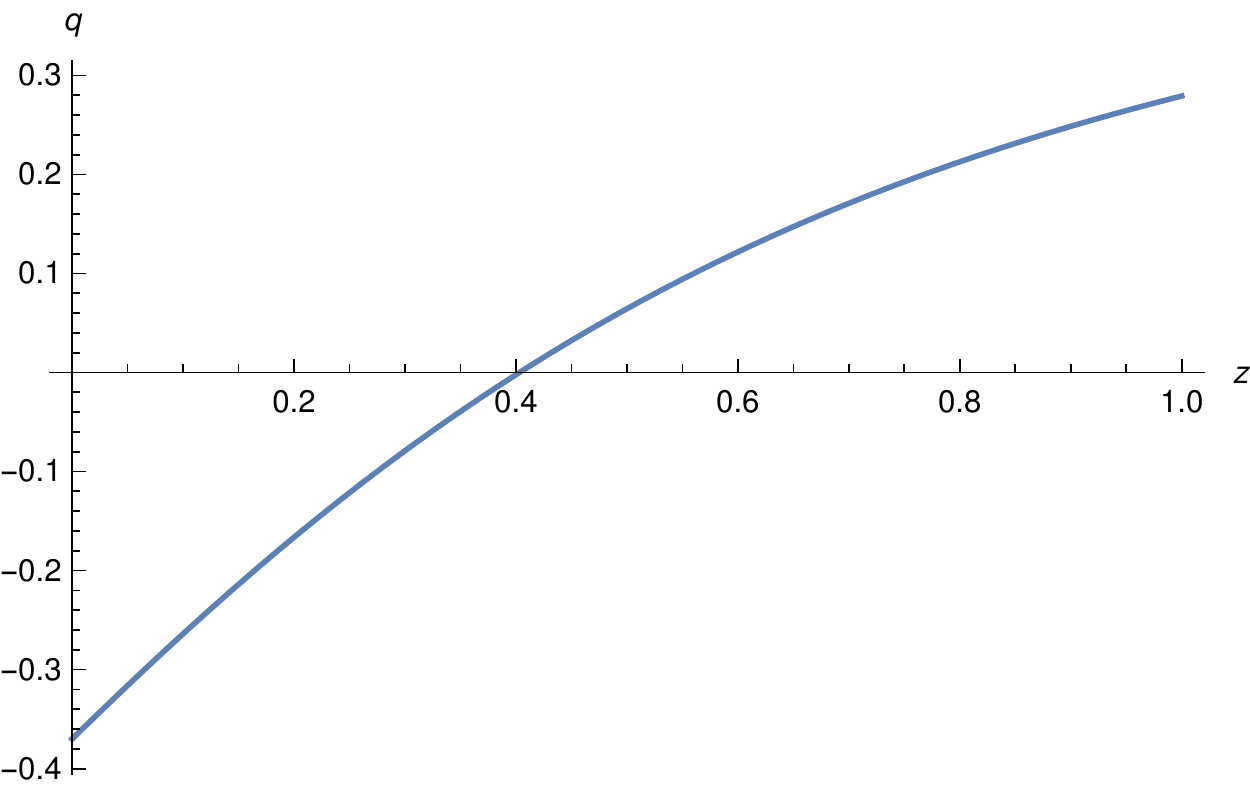}}
\caption{Plot of $q$ against $z$}
\end{figure}
Figure (1) shows the behaviour of $q$ against $z$. At $z=-1+2^{1/3}\sinh^{2/3}(1)\simeq 0.403$, expansion of the Universe transits from the deceleration to the acceleration phase. \\

With this model, the thermodynamic behaviour will be investigated.

\section{Thermodynamic Stability and Phase transition:}
  For thermodynamic stability, entropy of the fluid inside the horizon has to be maximized. In terms of the Hessian matrix of entropy, this can be realized as follows\cite{callen1998thermodynamics,kubo1968thermodynamics,carter2000classical,Muller1985}, all 
  the $k^{\text{th}}$ order principle minors of the matrix are $\leq$ 0 if $k$ is odd and $\geq$ 0 if $k$ is even. Hessian matrix $W$ of $S_{\mbox{in}} $ is

\begin{equation}
W=
\begin{bmatrix}
\frac{\partial^2{S_{\mbox{in}}}}{\partial U^2} & \frac{\partial^2{S_{\mbox{in}}}}{\partial U\partial V}\\ \vspace{0.005mm}\\
\frac{\partial^2{S_{\mbox{in}}}}{\partial V\partial U} & \frac{\partial^2{S_{\mbox{in}}}}{\partial V^2}
\end{bmatrix}.
\end{equation}

Therefore the thermodynamic stability requires that the conditions
\begin{align} 
(i)\frac{\partial^2{S_{\mbox{in}}}}{\partial U^2}\leq 0, \label{condi1}\\
(ii)\frac{\partial^2{S_{\mbox{in}}}}{\partial U^2}\frac{\partial^2{S_{\mbox{in}}}}{\partial V^2}- \left(\frac{\partial^2{S_{\mbox{in}}}}{\partial U\partial V}\right)^2\geq 0, \label{condi2}
\end{align}

are satisfied together. \\

Now, 

\begin{equation}\label{fc}
\frac{\partial^2{S_{\mbox{in}}}}{\partial U^2} = -\frac{1}{T^2C_V},
\end{equation}
and 
\begin{equation}\label{sc}
 \frac{\partial^2{S_{\mbox{in}}}}{\partial U^2}\frac{\partial^2{S_{\mbox{in}}}}{\partial V^2}-\left(\frac{\partial^2{S_{\mbox{in}}}}{\partial U\partial V}\right)^2=\frac{1}{C_VT^3V\beta_T} = \alpha.
\end{equation} The second expression is denoted by $\alpha$ for the sake of brevity.  Here $T$ is the temperature, $C_V$ is heat capacity at constant volume 
and $\beta_T$ is the isothermal compressibility. \\

Heat capacity at constant volume ($C_V$) and that at constant pressure ($C_P$)  of the fluid are defined respectively as,
\begin{equation}\label{cv}
C_V= T\left(\frac{\partial S_{\mbox{in}}}{\partial T}\right)_V,
\end{equation}
and
\begin{equation}\label{cp}
C_P= T\left(\frac{\partial S_{\mbox{in}}}{\partial T}\right)_P.
\end{equation}
Isothermal compressibility is defined as, 
\begin{equation}\label{bet}
\beta_T =-\frac{1}{V}\left(\frac{\partial V }{\partial P}\right)_T.
\end{equation}
Using equation \eqref{Gibbs}, one can calculate the heat capacities and isothermal compressibility for the matter inside the event horizon.
  
 \begin{align}\label{cv-expr}
 C_V &= V\left(\frac{\partial \rho}{\partial T}\right)_V \nonumber \\
 &= 32\pi^2\frac{\dot{H}}{2H^2\dot{H}+H\ddot{H}-\dot{H}^2}\nonumber\\
&= \frac{144\pi^2}{-2+(1+z)^3\csch^2(1)},
 \end{align}
 
 \begin{align}
 C_P &=  V\left(\frac{\partial \rho}{\partial T}\right)_P+(\rho+P)\left(\frac{\partial V}{\partial T}\right)_P \nonumber \\
 &= 32\pi^2\frac{H^2\dot{H}+\dot{H^2}}{H^2(2H^2\dot{H}+H\ddot{H}-\dot{H}^2)} \nonumber\\
 &= -72\pi^2 \frac{\sinh^2(1)}{(1+z)^3+\sinh^2(1)},
 \end{align}
 
\begin{align}\label{beta-expr}
 \beta_T &= \frac{3\dot{H}(2H^2+\dot{H})}{2(H^2+\dot{H})(2H^2\dot{H}+H\ddot{H}-\dot{H}^2)} \nonumber\\
 &=-\frac{27}{4}\frac{4+(1+z)^3\csch^2(1)}{\left(-2+(1+z)^3\csch^2(1)\right)}.
 \end{align}

 \begin{figure}
 \centering
\boxed{\includegraphics[scale=0.6]{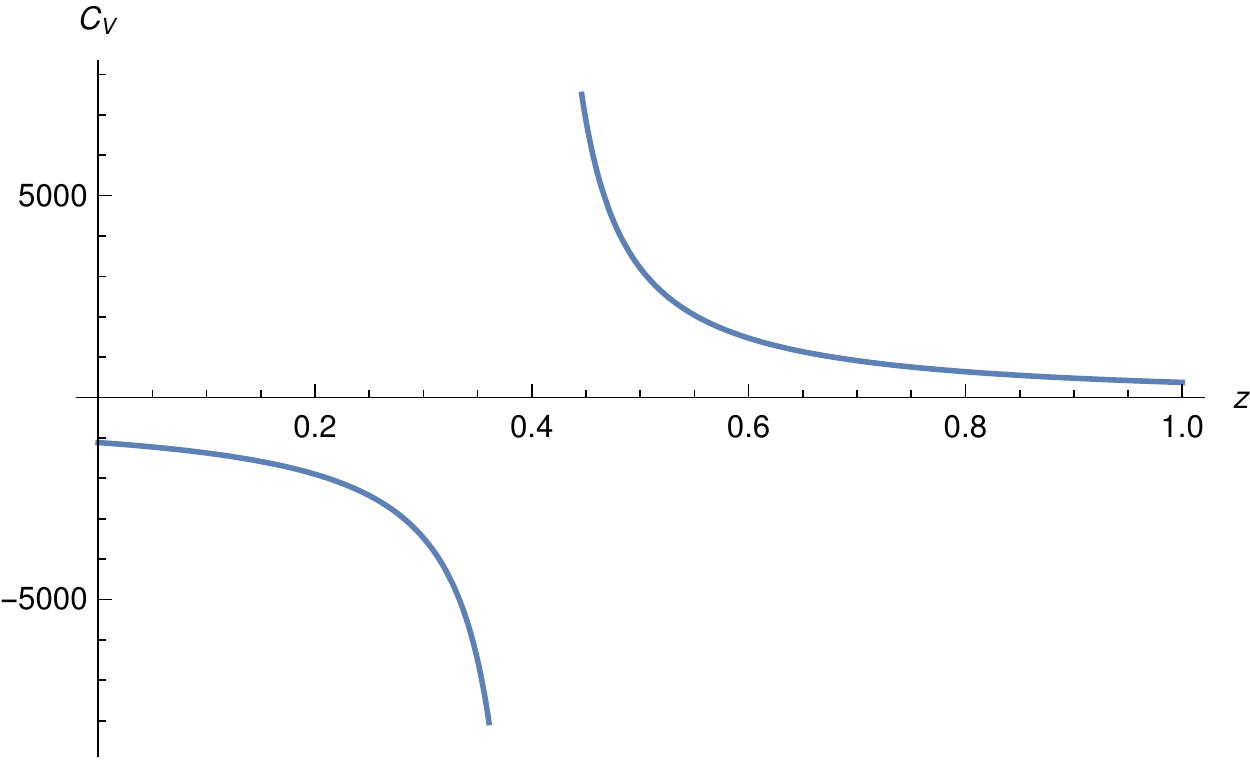}}
\caption{Plot of $C_V$ against $z$}\label{cvfig}
\end{figure}.

\begin{figure}
\centering
\boxed{\includegraphics[scale=0.6]{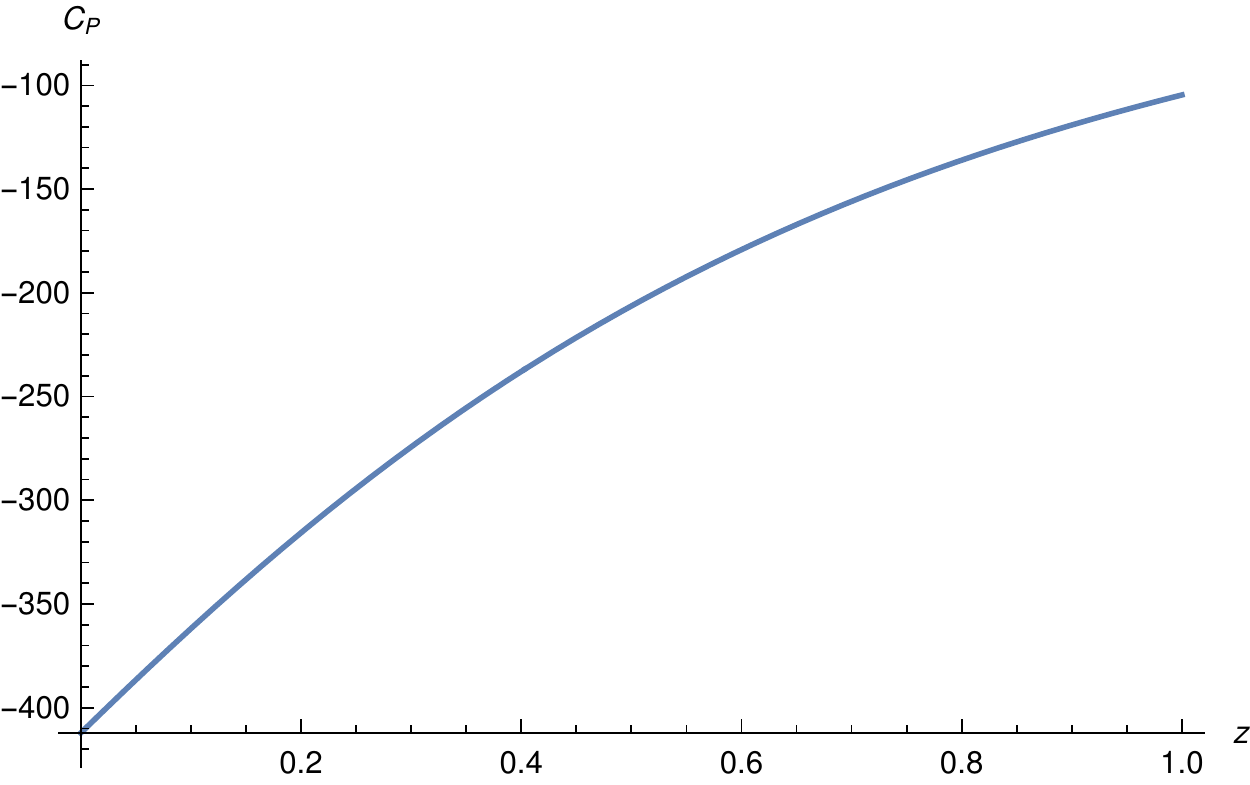}}
\centering\caption{Plot of $C_P$ against $z$}\label{cpfig}
\end{figure}

\begin{figure}
\centering
 \boxed{\includegraphics[scale=0.6]{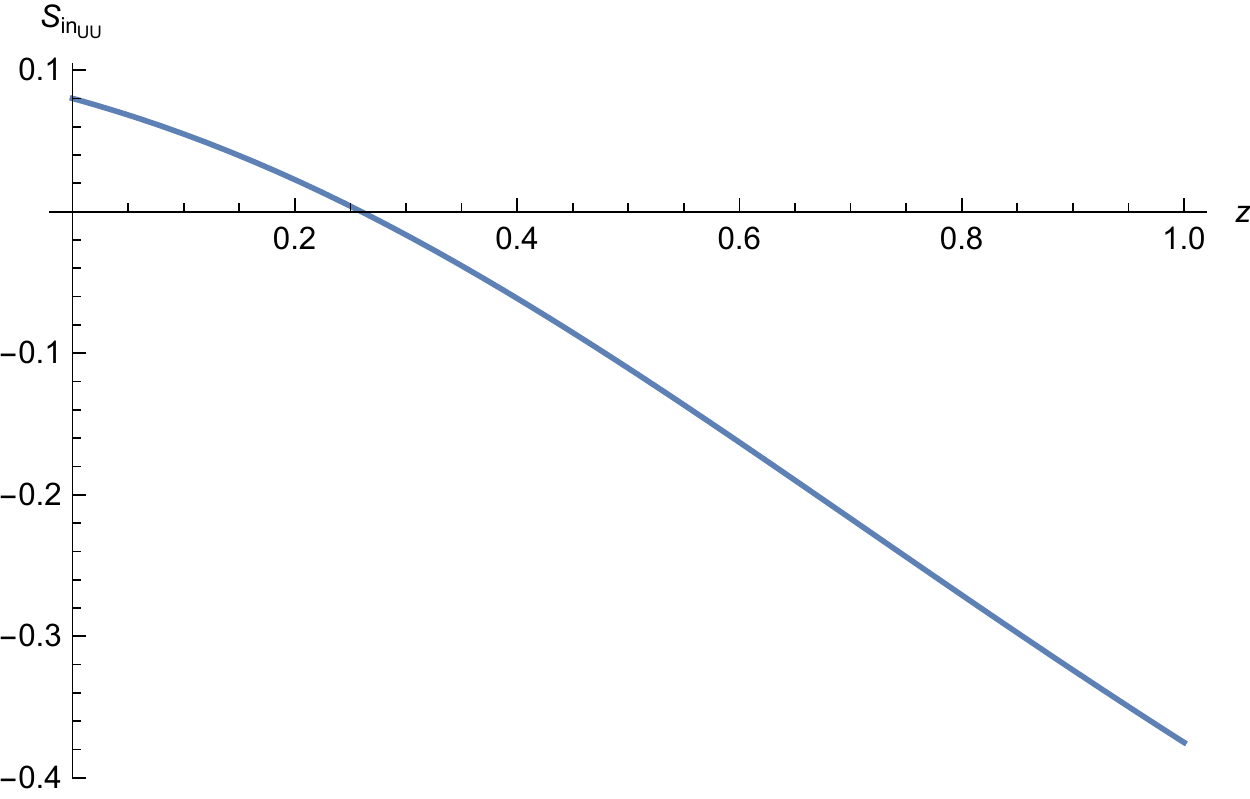}}
\caption{Plot of ${S_{in}}_{UU}$  against $z$}\label{cond1}
 \end{figure}

\begin{figure}
 \centering
 \boxed{\includegraphics[scale=0.6]{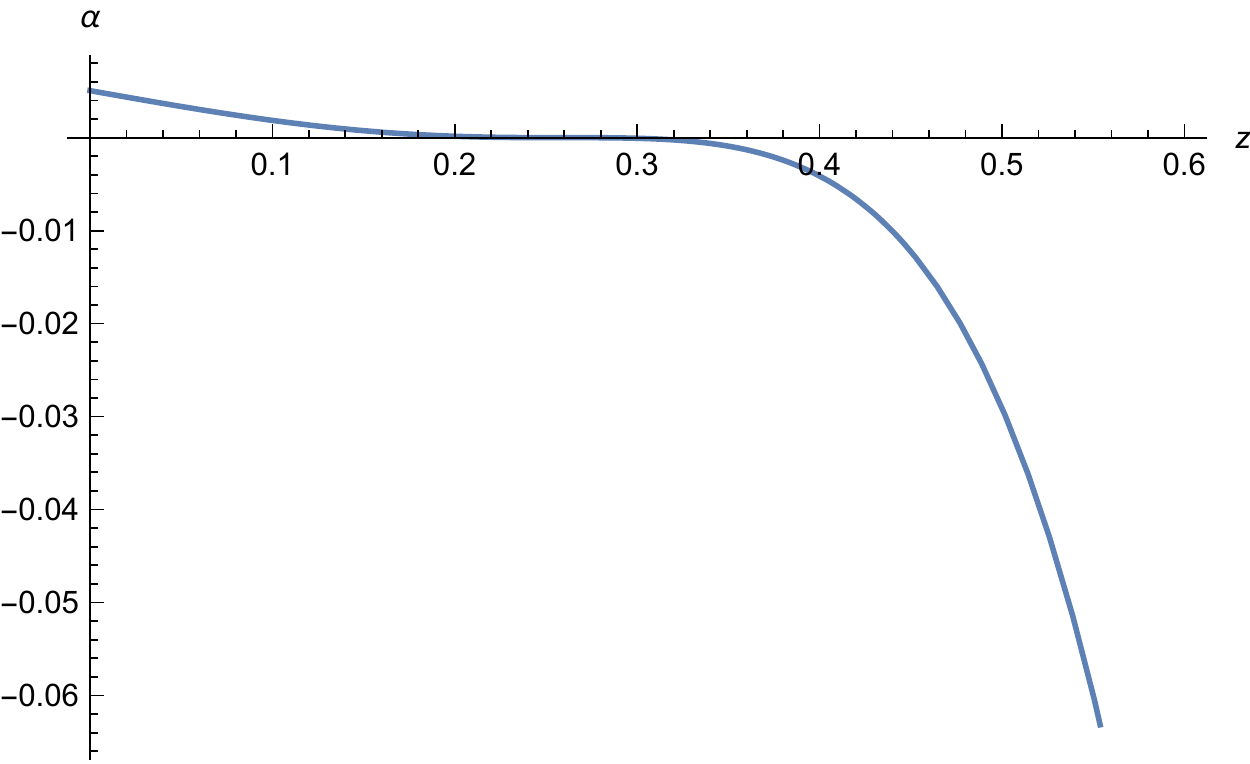}}
 \caption{Plot of $\alpha$ against $z$}\label{cond2}
 \end{figure}

Figures(\ref{cvfig}), (\ref{cpfig}) describes the behaviour of $C_V$ and $C_P$  respectively against the redshift $Z$ for low $z$ ($0\leq z \leq 1$). Using the expressions for $C_V$ and $\beta_T$ from equations  
\eqref{cv-expr} and \eqref{beta-expr} in equations \eqref{fc} and \eqref{sc}, one can plot $\frac{\partial^2{S_{\mbox{in}}}}{\partial U^2}$ and $\alpha$. These are shown in figures \eqref{cond1} and \eqref{cond2} respectively. In figure \eqref{cond1}, we have written ${S_{in}}_{UU}$ as legend along Y-axis and in the caption in place of $\frac{\partial^2{S_{\mbox{in}}}}{\partial U^2}$. It is clearly seen that the two 
conditions \eqref{condi1} and \eqref{condi2} are never satisfied together for the low redshift range ($0\leq z \leq 1$). Thus the model is not thermodynamically stable in the said redshift range. \\

The crucial observation that one can make from the figure \eqref{cvfig} is that $C_V$ has a discontinuity, in fact a divergence, at $z=-1+2^{1/3}\sinh^{2/3}(1)\simeq 0.403$, the value of $z$ where the Universe flips from the 
decelerated to the accelerated phase of expansion! So, the transition from the decelerated to the accelerated state of expansion is actually a thermodynamic phase transition. It has been checked that the entropy $S$ does not have any such 
discontinuity in the mentioned redshift range, the discontinuity is rather in $C_V$. Thus the phase transition is definitely a second order phase transition. \\

It deserves mention that $C_V$ is negative for the present Universe, $z>-1+2^{1/3}\sinh^{2/3}(1)\simeq 0.403$. However, a negative heat capacity is not at all a surprise in gravitational systems (for a review, see\cite{padmanabhan1990statistical}). In fact,
Luongo and Quevedo\cite{luongo2014cosmographic} arrived a strong result that for a currently accelerating Universe, $C_V$ is required to be negative. \\

Using equation \eqref{q} in \eqref{cv-expr}, one can obtain $C_V$ in terms of $q$ as
\begin{equation}\label{cvq}
 C_V=24\pi^2\frac{1-2q}{q}.
\end{equation}

So it is clearly seen that the discontinuity in $C_V$ results from $q$ appearing in the denominator with an exponent $+1$. Thus $q$ serves as the order parameter and the discontinuity is of order unity.

\section{Conclusion:}
The thermodynamical stability analysis is done for a model which mimics the $\Lambda$CDM model for the present Universe. As the horizon is evolving, the Hayward-Kodama temperature is considered as the horizon temperature. The thermal stability of an equilibrium thermodynamic system  is ensured   by demanding a positive thermal capacity and compressibility of the system and the same holds true for the  matter content  in a cosmological system(see for example the recent work of Luciano\cite{luciano}). In the present case, $C_V$ comes out to be negative. Thus the model is expected to have a thermodynamical instability. \\

The far reaching result obtained is that the matter content undergoes a phase transition as the Universe flips from the decelerated to the accelerated state of expansion. The phase transition is manifestly a second order one, as the 
discontinuity is in $C_V$. The deceleration parameter $q$ plays the role of the order parameter. One disclaimer is that it has not been the attempt to fit in the observational value of $z$ at $q=0$. The investigation was really that 
of the qualitative thermodynamic nature of the signature flip in $q$. The reason for earlier investigations being unable to find the second order phase transition at the commencement of the accelerated phase of expansion of the Universe is 
perhaps because of the use of the Hawking temperature of the horizon and thus ignoring the fact that the apparent horizon is evolving. \\

  It is generally believed that the early universe witnessed at least two phase transitions, one is the electroweak phase transition at the energy scale of $100$ GeV (temperature $\sim 10^{15}$ K), the other being the QCD 
transition at a temperature of $10^{12}$ K\cite{anupam}. There are speculations about some other phase transitions as well\cite{boyanovsky}. The idea of thermal or non-thermal phase transitions in cosmic matter all began when 
particle physics theories started using the universe as the laboratory in the quest of high energies unattainable in the laboratories on the earth. The combination of gravity and particle physics could describe the 
structure of the universe quite well. This communication between gravity and particle physics, however, is possible after the Planck epoch of energy scale beyond $10^{19}$ GeV, for which a proper quantum theory of 
gravity is required, which is still elusive. For a comprehensive review on the essence of the phase transitions in cosmic matter, we refer to the classic work of Kibble\cite{kibble}. Although some indirect effects of these transitions 
in the observational quantities are expected\cite{boyanovsky} in the present epoch, there is not much indication of a thermal phase transition in the late universe itself. The present work brings out the possibility that the cosmic 
fluid in the late universe itself might have undergone a phase transition at the value of the redshift at which a decelerated expansion of the universe enters into an accelerated mode defying the attractive nature of normal matter and 
thus requiring the so called dark energy. As opposed to the first order transitions in the early universe\cite{hindmarsh}, this late time phase transition is strongly second order. Certainly this late time phase transition deserves more attention, as this may hold the key to uncover the mystery of the dark energy.

We have assumed here that the dark matter and the dark energy are in thermal equilibrium with the horizon. This may appear to be a bit contrived\cite{Pavon2007gt} as they may evolve independently. However, we have assumed only a composite fluid where various sectors are not distinguished and only the evolution history ($a\sim \sinh^{2/3}(t/t_0)$) matters.

\section{Acknowledgment}
TD wants to thank Council of Scientific \& Industrial Research, India (CSIR) for financial support.

\bibliography{biblio}

\end{document}